\documentclass[aps,english,prb,twocolumn,reprint,superscriptaddress]{revtex4}
\usepackage{graphicx}
\usepackage[left]{lineno}
\usepackage{gensymb}
\usepackage{amsmath}

\makeatletter
\def\p@figure{}
\makeatother

\usepackage[version=4]{mhchem}

\begin{document}

\title{Interface Modeling of Perovskite/Polymer Heterostructures for Enhanced Charge Transfer Efficiency in Hybrid Photovoltaic Materials}

\author{Somayyeh Alidoust}
\affiliation{Faculty of Engineering and Natural Sciences, Sabanci University, Istanbul, 34956 Turkey.}

\author{V. Ongun {\"O}z{\c{c}}elik} \email{ongun.ozcelik@sabanciuniv.edu}
\affiliation{Faculty of Engineering and Natural Sciences, Sabanci University, Istanbul, 34956 Turkey.}
\affiliation{Materials Science and Nano Engineering Program, Sabanci University, Istanbul, 34956 Turkey.}

\begin{abstract}
Perovskite solar cells (PSCs) based on methylammonium lead iodide ($\ce{MAPbI3}$) exhibit remarkable photovoltaic performance, where interface engineering with hole transport layers (HTLs) is crucial for optimizing charge transfer and device efficiency. In this work, we present a density functional theory (DFT) study of the $\ce{MAPbI3}$/poly(3-hexylthiophene) (P3HT) hybrid interface, focusing on the role of perovskite surface termination in determining interfacial stability and electronic structure. We model MAI- and PbI-terminated $\ce{MAPbI3}$ surfaces interfaced with P3HT and compare their interfacial electronic properties. Electronic structure calculations reveal distinct differences in orbital hybridization and band alignment: the MAI/m-P3HT interface exhibits weak coupling, whereas the PbI/m-P3HT interface shows stronger hybridization and enhanced charge transfer. Band alignment confirms type-II, hole-selective character in both cases, with more pronounced VBM adjustment for PbI. Charge difference maps, Bader analysis, and local density of states (LDOS) consistently indicate higher charge transfer and stronger electronic coupling for PbI termination. Electrostatic potential offsets and transport parameters further highlight termination-dependent differences, with lighter effective masses at PbI/m-P3HT and higher hole velocity at MAI/m-P3HT. These findings provide theoretical insight into interfacial charge transport mechanisms and offer guidelines for optimizing perovskite–organic hybrid solar cells.
\end{abstract}

\maketitle

\section{Introduction}

Heterostructures formed by interfacing organic semiconducting polymers with stable substrates have gained significant attention for flexible electronics and photovoltaic technologies \cite{bae2019integration, huynh2002hybrid, chang2010high, ozcelik2020hybrid}. The contact between the polymer and the substrate surface often triggers spontaneous interfacial charge transfer, leading to electron–hole separation and redistribution of carriers, which reshapes the interfacial band structure via hybridization-driven charge accumulation \cite{xiang2017ultrafast, braun2009energy}. The effectiveness of this charge transfer is highly sensitive to the atomic arrangement at the junction, interfacial electronic and chemical properties, polymer adsorption configuration, and hybridization strength, collectively setting the band alignment, carrier injection barriers, and contact resistance \cite{dag2008modeling, dag2010packing}. Among these heterostructures, perovskite based solar cells (PSCs) have emerged as a leading class of photovoltaic materials due to their high power conversion efficiencies (PCEs) (exceeding 27\%), mechanically flexible optoelectronic properties, and compatibility with low-temperature, solution-based processing, which contributes to reduced fabrication costs \cite{Grancini2019,Tyznik2021,nrel.gov}{.} For example, methylammonium lead iodide ($\ce{CH3NH3PbI3}$ or $\ce{MAPbI3}$), which has a band gap of 1.55 eV, was first employed in liquid dye-sensitized solar cells in 2009, sparking a rapid surge in scientific research aimed at discovering improved perovskite materials for solar cells\cite{Kojima2009}{.} Just like $\ce{MAPbI3}$, PSC materials typically adopt the $\ce{ABX3}$ perovskite structure, where A is a monovalent cation, B is a divalent metal cation, and X is a halide anion. The choice of each component directly affects the crystal stability, band gap, and overall optoelectronic performance of the perovskite. These characteristics make $\ce{ABX3}$ perovskites highly effective as an absorber layers in solar cell applications.

\begin{figure*}[htbp]
	\centering
	\includegraphics[ width=1\textwidth]{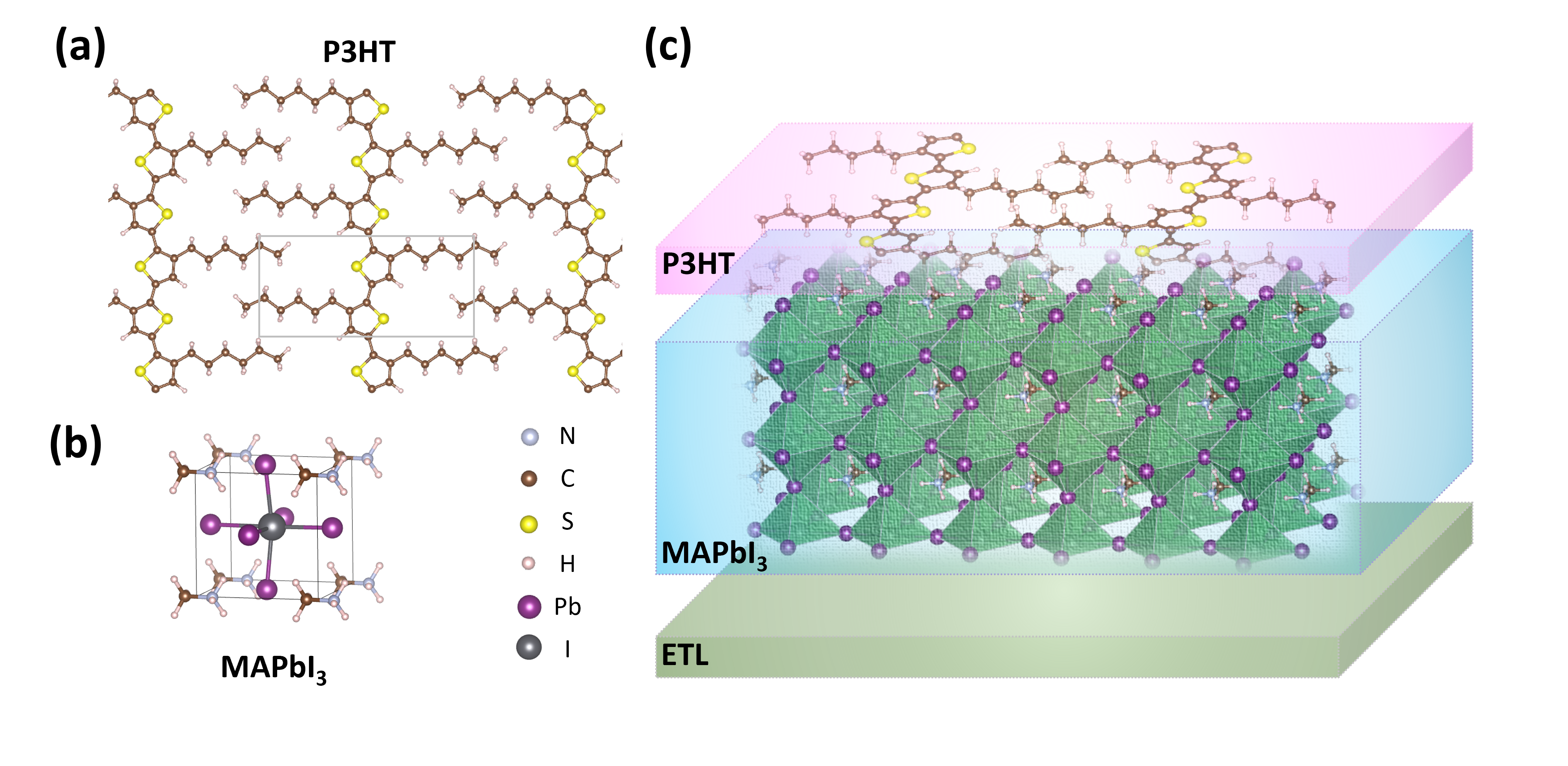}
	\caption{(a) 2D periodic structure of P3HT. (b) 3D crystal structure of MAPbI$_3$. (c) Schematic illustration of the MAPbI$_3$/P3HT interface used in this study.}
	\label{fig1}
\end{figure*}

In a typical PSC, the perovskite layer is sandwiched between electron and hole transport layers (ETL and HTL), which are crucial for efficient charge extraction. The performance of a PSC is heavily influenced by the interfaces between the perovskite layer and both the ETL and HTL, where energy level alignment, charge carrier mobility, and interfacial recombination play key roles in determining device efficiency and stability. Numerous studies on PSC devices have appeared in the literature, aiming to enhance power conversion efficiency, stability, and scalability, where these efforts have mainly focused on optimizing the perovskite absorber layer and improving the charge transport layers, including both the HTL and ETL\cite{C6CP00721J,REHMAN2025112487,Hossain2025,Alidoust2024,Li2021,Vo2024,Chen2022,Dkhili2022}{.} For the perovskite layer, strategies such as MA, Pb, and I-site substitution have been widely explored. For example, formamidinium lead iodide ($\ce{FAPbI3}$) has been developed to MA for improved thermal stability\cite{C6CP00721J}{.} Similarly, potassium germanium chloride ($\ce{KGeCl3}$) represents an approach to substitute lead with less toxic metals like germanium, while also demonstrating promising photovoltaic performance with relatively high PCE\cite{REHMAN2025112487}{.} Additionally, mixed-cation and mixed-halide perovskites, such as $\ce{(MA, FA)PbI3}$\cite{CHOI201956} or $\ce{(FA, Cs)Pb(I, Br)3}$\cite{Heinze2024}{,} have been investigated to enhance phase stability and tune optoelectronic properties. For the ETL, various materials have been utilized, including inorganic semiconductors such as ZnO and TiO$_2$, along with organic compounds like [6,6]-phenyl-C$_{61}$-butyric acid methyl ester (PCBM) and C$_{60}$\cite{Dkhili2022}{.} The HTLs can enhance hole extraction and affect the open-circuit voltage $\ce{V_{oc}}$ by increasing hole concentration at the contact since  $\ce{V_{oc}}$ of PSCs is governed by the quasi-Fermi level difference at the interfaces\cite{Westbrook2018}. Organic HTLs such as poly(3,4-ethylenedioxythiophene):poly(styrenesulfonate) (PEDOT:PSS) and Spiro-OMeTAD are common choices, while inorganic alternatives like CuI, NiO, and Cu$_2$O-doped Spiro-OMeTAD have also been investigated to enhance hole mobility and device longevity \cite{Li2021,Vo2024}{.}

Poly(3-hexylthiophene) (P3HT) is a benchmark semiconducting conjugated polymer extensively employed in organic photovoltaic and solar cell technologies due to its favorable charge-transport characteristics, ease of solution-based processing, and adjustable optoelectronic properties \cite{liu2009photovoltaic, fratini2020charge, lucke2015p3htdefects, liu2011nanofibril, rudyak2020stacking, mardi2020thermoelectric}. Due to the high cost of conventional HTLs like Spiro-OMeTAD, P3HT has been explored as a cost-effective alternative in PSCs\cite{Baraneedharan2024-nk,XIE2022128072,Xu2022}{.}  Recent studies have shown that modifications, blends, and interface engineering with the perovskite layer can enhance its charge transport properties\cite{Baraneedharan2024-nk,XIE2022128072,Xu2022}. An additive-assisted strategy using P3HT in the perovskite precursor solution enhances film crystallinity, reduces defects, and creates a favorable perovskite/P3HT heterojunction, leading to improved hole extraction and increased PCE, from 12.72\% to 15.57\% in rigid and 11.81\% to 13.54\% in flexible devices\cite{XIE2022128072}{.} Xu et al.\cite{Xu2022} proposed constructing a molecular bridge at the P3HT/perovskite interface, which significantly enhances both the efficiency and stability of the resulting solar cells. This approach facilitates improved hole extraction, suppresses interfacial recombination, and passivates interfacial defects, leading to devices that demonstrate competitive PCEs alongside enhanced operational stability under ambient conditions\cite{Xu2022}{.}

In this study, we focus on the interfacial optoelectronic properties of the MAPbI$_3$/P3HT system, shown in Figure \ref{fig1}, using theoretical methods to better understand charge transfer behavior and band alignment at the interface. We select MAPbI$_3$ as the perovskite material due to its well-established role as a benchmark in perovskite solar cell research. Its high PCE and extensive prior characterization make it an ideal reference point for studying interfacial properties and charge transport mechanisms. This approach can be extended to other perovskite compositions with reduced toxicity through B-site substitution, and can further be used to investigate how variations in the A-, B- and X-sites components influence interfacial behavior. We model the interfaces with different surface terminations, specifically MAI- and PbI$_2$-terminated surfaces, to examine how termination affects the electronic coupling at the junction. This work offers a theoretical framework which has not yet been systematically explored for this type of interface. Using density functional theory (DFT), we compute band structures, band alignments, and charge transport characteristics, providing a comparative analysis between different terminations. Our results reveal clear interactions at the interface, including hybridization of states, which impact charge transfer dynamics and alignment. These findings offer valuable insight into the design of perovskite–organic hybrid solar cells and guide future efforts toward optimizing interfacial charge separation and transport.

\section{Methods}
All first-principles calculations were performed using the Vienna Ab initio Simulation Package (VASP)\cite{PhysRevB.54.11169}{.} The electronic structure was treated within the generalized gradient approximation using the Perdew–Burke–Ernzerhof (PBE)\cite{PhysRevLett.77.3865} exchange–correlation functional. Long-range dispersion interactions were included through the DFT-D3 correction\cite{10.1063/1.5023802}{.} Projector-augmented-wave (PAW) pseudopotentials\cite{PhysRevB.50.17953} were employed to describe the electron–ion interactions. For each structure, full geometry optimizations were performed by relaxing all atomic positions and the cell volume, while keeping the lattice angles fixed to preserve the overall cell shape. The plane-wave basis set was expanded using an energy cutoff of 500 eV. A k-point mesh with approximately 40 points along each periodic direction was used to ensure convergence. The electronic self-consistency criterion was set to 10$^{-5}$ eV, and ionic relaxations were continued until the residual forces on all atoms were below 10$^{-3}$ eV/Å. Spin–orbit coupling (SOC) and hybrid exchange–correlation functionals were not applied. Although we are aware that the PBE functional underestimates the band gap, the very large number of atoms in our interface models made the inclusion of SOC or hybrid functional calculations computationally prohibitive. A vacuum spacing of 30 Å was introduced along the non-periodic direction to eliminate spurious interactions between periodic images, and dipole corrections were applied along the non-periodic direction to eliminate artificial electrostatic interactions in the systems. All static electronic-structure calculations, including the projected density of states, charge-density differences, and Bader charge analysis, were performed using VASP in conjunction with VASPKIT\cite{VASPKIT} and PyProcar \cite{HERATH2020107080}{.}

\section{Results and discussion}
\begin{figure*}
	\centering
	\includegraphics[width=0.9\textwidth]{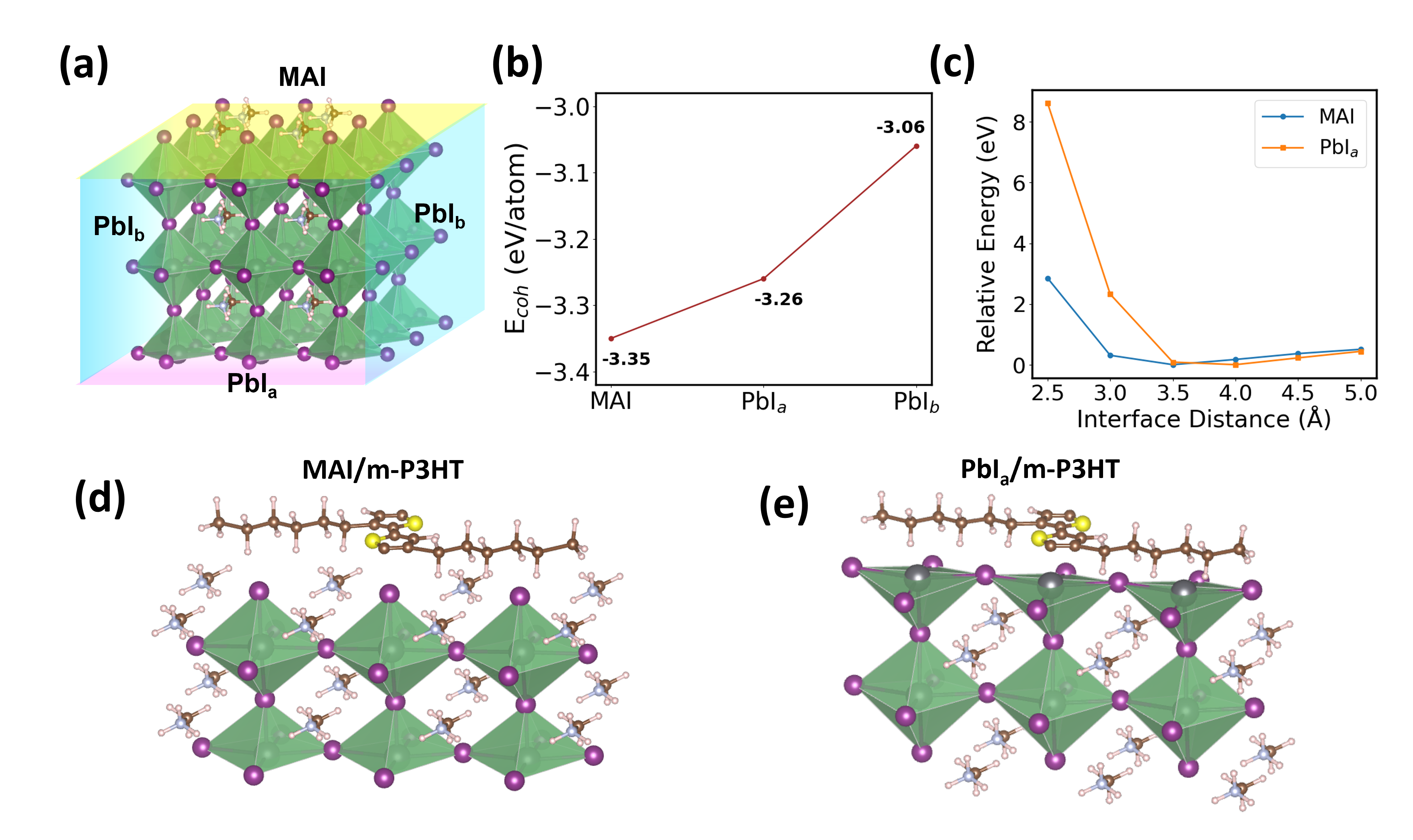}
	\caption{(a) MAI, PbI$_a$, and PbI$_b$ terminations of MAPbI$_3$ used for the P3HT interface study.
		(b) Cohesive energy per atom for each termination; lower values indicate greater stability.
		(c) Relative total energy versus distance between perovskite surface and m-P3HT; minima mark optimal interface separations.
		(d,e) Optimized structures of MAPbI$_3$/m-P3HT interfaces for MAI- and PbI$_a$-terminations, respectively..}
	\label{Fig2}
\end{figure*}

To support the analysis of interface stability, three terminations of pseudocubic triclinic $P1$ structure of MAPbI$3$\cite{WMD_group2014CH3NH3PbI3} perovskite were considered, as shown in Figure~\ref{Fig2}a, for the calculation of cohesive energy ($E_{coh}$) and the selection of suitable terminations for subsequent interface modeling. These were selected from among the variety of surface terminations considered in previous studies\cite{Kasi2018,Wang2015,Haruyama2016}{.} The terminations correspond to different surface layers: one terminated by methylammonium and iodide ions, referred to as MAI, and two terminated by lead and iodide ions with distinct atomic arrangements, labeled PbI$_a$ (PbI$_5$) and PbI$_b$ (PbI$_6$). Notably, Matta et al.\cite{Kasi2018} also employed the MAI and PbI$_a$ surfaces in their investigation of the perovskite/HTL interface. Here, we revisit these specific terminations to provide a comparative assessment based on cohesive energy.

The cohesive energy for each slab was computed using the following formula:

\begin{equation}
	E_{\text{coh}} = \frac{E_{\text{slab}} - \sum_i n_i E_i^{\text{atom}}}{N}
\end{equation}

where $E_{slab}$ is the total energy of the slab, $E_{i}^{atom}$ is the reference energy of each isolated atom, $n_{i}$  is the number of atoms of type i, and N is the total number of atoms in the slab. Figure \ref{Fig2}b shows the calculated cohesive energy per atom for different surface terminations of the perovskite slab. Lower negative values (higher magnitude) indicate higher cohesive energy and greater thermodynamic stability. Among the considered terminations, MAI (3.35 eV/atom) and PbI$_a$ (3.26 eV/atom) exhibit slightly higher cohesive energies compared to PbI$_b$ (3.03 eV/atom), suggesting a marginally higher stability. Although the energy difference is not substantial, we focus on the MAI and PbI$_a$ terminations in the following analysis to reduce computational complexity and to prioritize the most stable configurations.

To form a commensurate heterostructure, three formula units of the selected MAPbI$_3$ termination were used as the perovskite slab, together with one formula unit of P3HT. As explained in detail in the Supporting Information (SI), P3HT was modified to match the dimensions of the three-unit MAPbI$_3$ layer. In the modified P3HT (m-P3HT), bond lengths and angles were altered compared to the pristine form. An initial vertical separation of 3.0 Å was introduced between the two materials prior to relaxation. The interface was then subjected to vertical distance optimization to determine the most favorable interlayer separation. Specifically, the distance between the m-P3HT layer and the perovskite surface was systematically varied from 2.5 Å to 5.0 Å . As shown in Figure \ref{Fig2}c, the total energy reached a minimum at a vertical separation of 3.5 Å, indicating the optimal distance for interfacial interaction. The optimized structures of the two resulting interfaces (MAI- and PbI$_a$-terminated) are shown in Figure \ref{Fig2}d and \ref{Fig2}e. An equilibrium distance of approximately 3.5 Å typically suggests a weak non-covalent interaction—most likely dominated by van der Waals forces—rather than strong chemical bonding, which usually occurs at shorter distances.

\begin{figure*}
	\centering
	\includegraphics[width=1\textwidth]{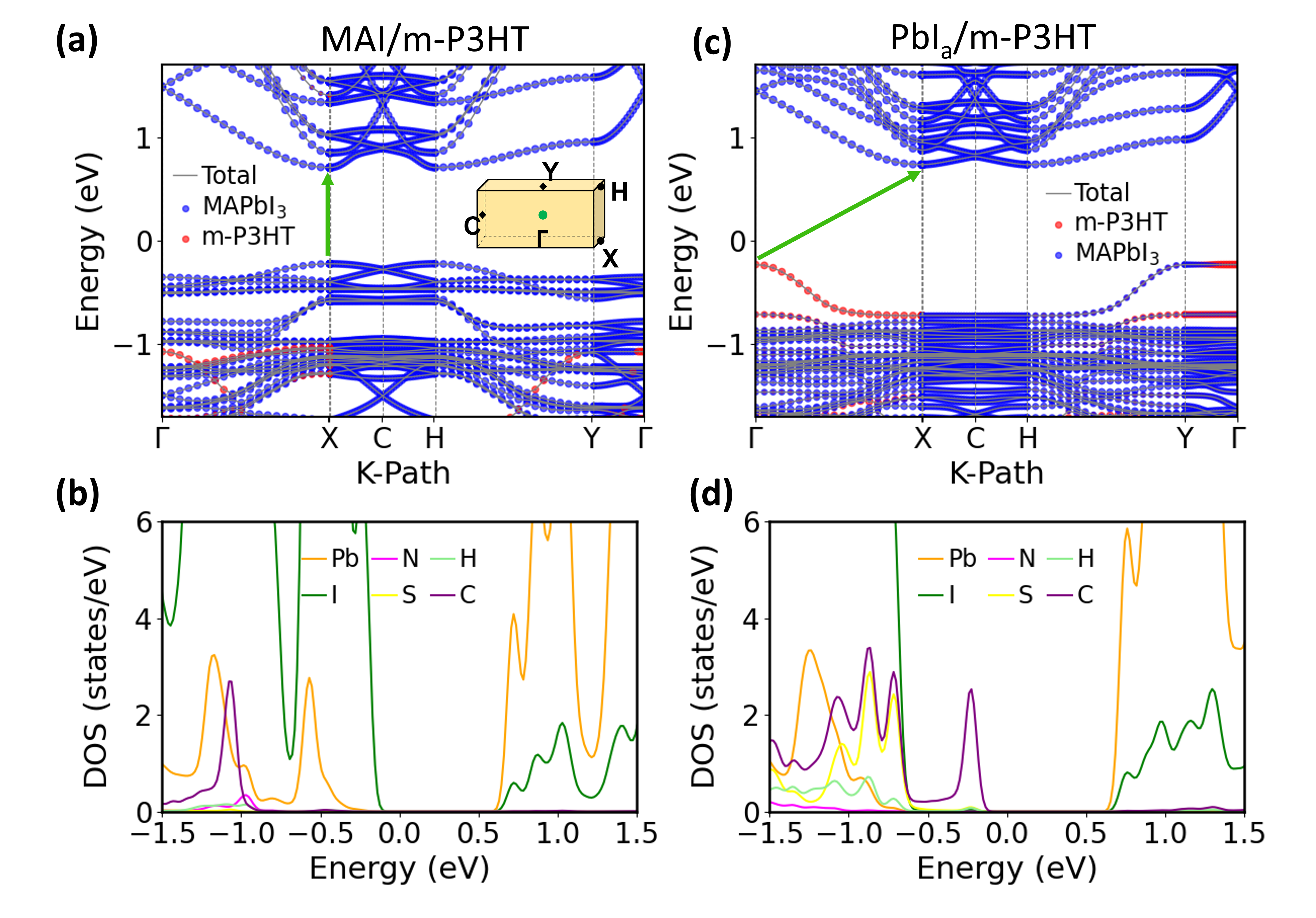}
	\caption{(a, b) Projected band structure and density of states of the MAI/m-P3HT interface (band gap = 0.93 eV), and (c, d) those of the PbI$_a$/m-P3HT interface (band gap = 0.97 eV), illustrating the electronic contributions from each component.}
	\label{Fig3}
\end{figure*}

To further investigate the electronic interactions at the interface, we analyzed the projected band structure of the MAI/m-P3HT and PbI$_a$/m-P3HT systems. Figure \ref{Fig3} reveals distinct contributions to the VBM and conduction CBM in the two interfaces. In the MAI/m-P3HT configuration, only the perovskite contributes near the band edges specifically, the VBM is primarily composed of iodine $p$-orbitals, while the CBM is dominated by lead $p$-orbitals. This indicates minimal interaction or hybridization between m-P3HT and the perovskite states near the Fermi level. In contrast, the PbI$_a$/m-P3HT interface exhibits noticeable orbital contributions from both materials. The VBM includes significant input from the carbon p-orbitals of the m-P3HT layer, with additional contribution from the iodine p-orbitals of the perovskite at slightly lower energy levels, while the CBM retains iodine character from the perovskite. This suggests stronger interfacial coupling in the PbI$_a$-terminated system, potentially facilitating improved charge transfer across the interface.

\begin{figure*}
	\centering
	\includegraphics[width=1\textwidth]{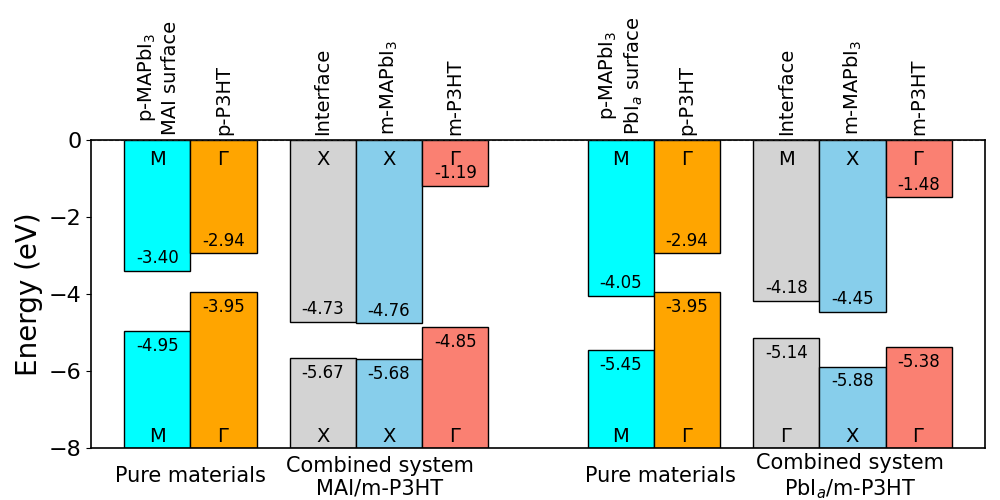}
	\caption{Band alignment of the MAI/m-P3HT and PbI$_a$/m-P3HT interfaces, both exhibiting type-II alignment. Gray bars represent the interface band edges, while skyblue and salmon bars indicate the modified perovskite (m-MAPbI$_3$) and modified P3HT (m-P3HT) band edges at the interface, respectively. Cyan and orange bars correspond to the pure perovskite (p-MAPbI$_3$ with MAI and PbI$_a$ terminations) and pure P3HT (p-P3HT), included for comparison.}
	\label{Fig4}
\end{figure*}

Band alignment analysis of the MAI/m-P3HT and PbI$_a$/m-P3HT interfaces reveals notable differences in energy level alignment that influence charge transport across the interface, as shown in Figure \ref{Fig4}. Both interfaces exhibit type-II band alignment\cite{ozccelik2016band}  and are hole-selective. In both cases, the perovskite CBM lies far below the P3HT CBM ($\Delta E \geq 3$~eV), ensuring efficient electron blocking. For the MAI/m-P3HT interface, the bands are positioned deeper compared to the PbI$_a$/m-P3HT case, and the interface VBM and CBM remain almost unchanged with respect to the modified perovskite, consistent with the band structure results. In contrast, at the PbI$_a$/m-P3HT interface, the VBM shows stronger adjustment and appears influenced by both the perovskite and P3HT states, leading to a more noticeable modification of the interface levels and efficient hole extraction. In addition, the figure also includes the pristine band edges of the pure materials, which allows a direct comparison between the pristine and modified structures and highlights the interfacial effects.

\begin{figure*}
	\centering
	\includegraphics[width=\textwidth]{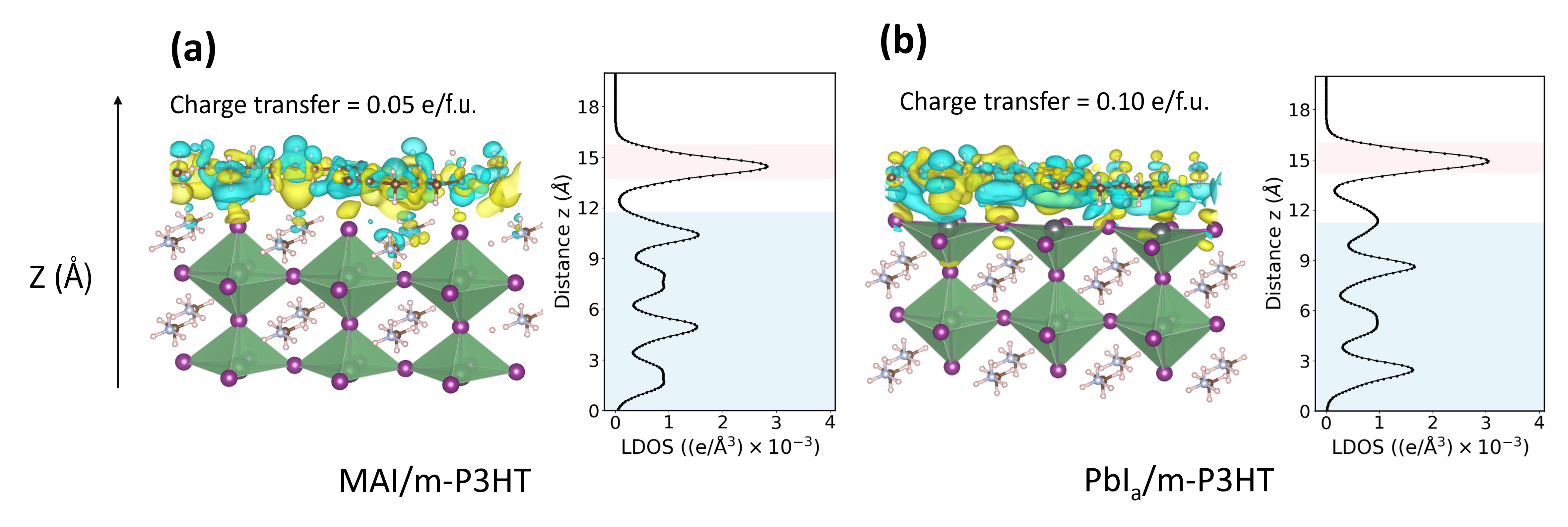}
	\caption{Charge density difference maps showing electron accumulation (yellow) and depletion (cyan) at the (a) MAI/m-P3HT and (b) PbI$_a$/m-P3HT interfaces. LDOS profiles, also shown here, reveal stronger interfacial coupling and charge delocalization in PbI$_a$ termination, supporting higher charge transfer and improved hole extraction.}
	\label{Fig5}
\end{figure*}

To understand the redistribution of electron density upon interface formation, charge difference analysis is carried out and calculated as:
\begin{equation}
	\Delta \rho = \rho_{\text{interface}} - \rho_{\text{perovskite}} - \rho_{\text{m-P3HT}}
	\label{eq:charge_diff}
\end{equation}

where $\rho_{interface}$ is the electron density of the combined system, and $\rho_{perovskite}$ and $\rho_{m-P3HT}$ are the electron densities of the isolated components in the same geometry. This difference highlights regions of charge accumulation and depletion at the interface. Figure \ref{Fig5} shows the charge difference analysis, along with Bader charge results that quantitatively partition the electron density among atoms to estimate net charge transfer and reveal the extent of charge rearrangement at the interface. The Bader charges indicate a slightly higher charge transfer at the PbI$_a$/m-P3HT interface (0.10 e/f.u. or $6.88\times10^{-4}$ e/\AA{}) compared to the MAI/m-P3HT interface (0.05 e/f.u. or $3.45\times10^{-4}$e/\AA{}), consistent with moderately stronger interfacial electronic interactions in the PbI$_a$-terminated system.

In addition to the charge difference and Bader charge analyses, the local density of states (LDOS) profiles further elucidate the spatial distribution of electronic states along the out-of-plane direction (z-axis). As shown in Figure~\ref{Fig5}, the LDOS at the MAI/m-P3HT interface is mainly localized within the perovskite slab, with only a weak extension into the molecular region. In contrast, the PbI$_a$/m-P3HT interface exhibits a more pronounced LDOS intensity at the boundary and a stronger extension toward the m-P3HT side. This result supports the larger Bader charge transfer and indicates enhanced electronic coupling at the PbI$_a$-terminated interface.

\begin{figure*}
	\centering
	\includegraphics[width=1\textwidth]{Fig6.png}
	\caption{Planar-averaged electrostatic potential along the z-direction for the (a) MAI/m-P3HT and (b) PbI$_a$/m-P3HT interfaces, along with the corresponding potential energy differences at the perovskite/m-P3HT interfaces. Shaded regions in blue and red indicate the MAPbI$_3$ and P3HT slabs, respectively.}
	\label{Fig6}
\end{figure*}

To ensure reliable extraction of interfacial electronic properties, we analyzed the planar-averaged electrostatic potential along the $z$-direction. As shown in Fig.\ref{Fig6}, the potential profiles for MAI/m-P3HT and PbI$_a$/m-P3HT exhibit pronounced oscillations within the $z \approx 0$–16\AA{} range, reflecting the periodic arrangement of atomic layers in the MAPbI$_3$ slab and the m-P3HT layer. Upon reaching the interfacial region, the oscillations are smoothed, and distinct potential offsets emerge between the perovskite and m-P3HT domains. The extracted potential energy differences ($\Delta V$) indicate a larger offset for the MAI/m-P3HT interface ($-2.7$~eV) compared to the PbI$_a$/m-P3HT interface ($0.23$~eV). These values highlight the sensitivity of interfacial energetics to surface termination and provide direct insight into the band alignment and charge transfer characteristics at the perovskite/organic interface.

The charge carrier transport properties at the MAI/m-P3HT and PbI$_a$/m-P3HT interfaces were evaluated by analyzing the effective masses and Fermi velocities of electrons and holes. As shown in Table \ref{tab:mass_velocity_combined}, both interfaces exhibit comparable electron Fermi velocities, while the hole Fermi velocity is higher at the MAI/m-P3HT interface. The effective masses show a clear difference: carriers are heavier at the MAI/m-P3HT interface ($m_e^\ast=1.30$, $m_h^\ast=1.11$) compared to the PbI$_a$/m-P3HT interface ($m_e^\ast=0.73$, $m_h^\ast=0.31$), indicating more dispersive bands in the latter case, especially for holes. These results suggest distinct transport characteristics depending on the interface termination, with lighter effective masses at PbI$_a$/m-P3HT and higher hole velocity at MAI/m-P3HT, reflecting different electronic structure modifications at the two interfaces.

\begin{table}[htbp]
	\centering
	\caption{Effective masses ($m^*$, in units of $m_0$) and Fermi velocities ($v_F$, in units of $10^{6}$~m/s) for electrons and holes at the MAI and PbI$_a$ interfaces.}
	\begin{tabular}{ccccc}
		\hline
		Interface & $m^*_\text{electron}$ & $m^*_\text{hole}$ & $v_{F, \text{electron}}$ & $v_{F, \text{hole}}$ \\
		\hline
		MAI/m-P3HT     &  1.30  & 1.11   & 1.72 & 1.87 \\
		PbI$_a$/m-P3HT & 0.73 & 0.31 & 1.73 &  1.39\\
		\hline
	\end{tabular}
	\label{tab:mass_velocity_combined}
\end{table}

\section{Conclusion}
In this conclusion, here we investigated the interfacial electronic properties of MAPbI$_3$/P3HT heterojunctions with MAI- and PbI$_a$-terminated surfaces using first-principles calculations. Cohesive energy analysis confirmed the relative stability of MAI and PbI$_a$ terminations, while structural optimization revealed an equilibrium interface distance of ~3.5 Å, consistent with van der Waals interactions. Electronic structure calculations demonstrated that the MAI/m-P3HT interface exhibits weak coupling, whereas the PbI$_a$/m-P3HT interface shows stronger hybridization and enhanced charge transfer. Band alignment analysis indicated type-II, hole-selective behavior for both interfaces, with improved VBM adjustment at the PbI$_a$ termination. Charge density differences, Bader charges, and LDOS profiles further supported the stronger interfacial electronic interactions in PbI$_a$/m-P3HT. Finally, transport analysis revealed lighter effective masses at the PbI$_a$ interface and higher hole velocity at the MAI interface, reflecting complementary charge transport characteristics. Overall, this theoretical insight provides a foundation for rational interface design in PSCs employing P3HT HTLs, guiding future efforts to optimize charge extraction and device performance.

\section*{Acknowledgements}
We acknowledges funding from the Scientific and Technological Research Council of Turkey (TUBITAK) under project no 123F264.  Computational resources have been provided by TUBITAK ULAKBIM, High Performance and Grid Computing Center and by the National Center for High Performance Computing of Turkey (UHeM) under grant numbers 1013432022.

\bibliographystyle{unsrt}

\bibliography{references.bib}


\end{document}